\documentclass[aps,prl,superscriptaddress,twocolumn]{revtex4-2}
\usepackage{comment}
\usepackage[utf8]{inputenc}
\usepackage{amssymb}
\usepackage{amsmath}
\usepackage{amsfonts}
\usepackage{bm}
\usepackage{mathtools}
\usepackage{bbold}
\usepackage{upgreek}
\usepackage{xfrac}
\usepackage{soul}
\usepackage{bm}
\usepackage{xcolor}
\usepackage{url}
\usepackage{txfonts}
\usepackage{bbm}
\usepackage{dsfont}
\usepackage{physics}
\usepackage{svg}
\usepackage[nolist]{acronym}

\DeclareSymbolFont{matha}{OML}{txmi}{m}{it}
\DeclareMathSymbol{\varv}{\mathord}{matha}{118}

\usepackage[colorlinks=true,citecolor=blue]{hyperref}
\hypersetup{colorlinks=true,citecolor=blue,linkcolor=blue,urlcolor=blue}
\usepackage{url}

\begin{document}

\title{Addressing the Readout Problem in Quantum Differential Equation Algorithms\\ with Quantum Scientific Machine Learning}
%\title{Quantum Differential Equation Solvers as Big Quantum Data Generators?}

\author{Chelsea A. Williams}
\affiliation{Department of Physics and Astronomy, University of Exeter, Stocker Road, Exeter EX4 4QL, United Kingdom}
\affiliation{PASQAL, 7 Rue Léonard de Vinci, 91300 Massy, France}
\author{Stefano Scali}
\affiliation{Department of Physics and Astronomy, University of Exeter, Stocker Road, Exeter EX4 4QL, United Kingdom}
\author{Antonio A. Gentile}
\affiliation{PASQAL, 7 Rue Léonard de Vinci, 91300 Massy, France}
\author{Daniel Berger}
\affiliation{Siemens AG, Gleiwitzer Str. 555, 90475 N\"urnberg, Germany}
\author{Oleksandr Kyriienko}
%\email{o.kyriienko@exeter.ac.uk}
\affiliation{Department of Physics and Astronomy, University of Exeter, Stocker Road, Exeter EX4 4QL, United Kingdom}
\affiliation{PASQAL, 7 Rue Léonard de Vinci, 91300 Massy, France}
\date{\today}

\begin{abstract}
Quantum differential equation solvers aim to prepare solutions as $n$-qubit quantum states over a fine grid of $O(2^n)$ points, surpassing the linear scaling of classical solvers. However, unlike classically stored vectors of solutions, the readout of exact quantum states poses a bottleneck due to the complexity of tomography. Here, we show that the readout problem can be addressed with quantum learning tools where we focus on distilling the relevant features.
% Treating outputs of quantum differential equation solvers as quantum data, we demonstrate that low-dimensional output can be extracted using the measurement operator adaptation for detecting relevant features.
Treating outputs of quantum differential equation solvers as quantum data, we demonstrate that low-dimensional output can be extracted using a measurement operator adapted to detect relevant features.
We apply this quantum scientific machine learning approach to classify solutions for shock wave detection and turbulence modeling in scenarios where data samples come directly from quantum differential equation solvers. We show that the basis chosen for performing analysis greatly impacts classification accuracy. Our work opens up the area of research where quantum machine learning for quantum datasets is inherently required.
\end{abstract}

\maketitle

\begin{acronym}
  \acro{CFD}{computational fluid dynamics}
  \acro{CNN}{convolutional neural network}
  \acro{DQNN}{deep quantum neural network}
  \acro{ML}{machine learning}
  \acro{PDE}{partial differential equation}
  \acro{QCNN}{quantum convolutional neural network}
  \acro{QFT}{quantum Fourier transform}
  \acro{QML}{quantum machine learning}
  \acro{QNN}{quantum neural network}
  \acro{QuaSciML}{quantum scientific machine learning}
  \acro{SciML}{scientific machine learning}
  \acro{QC}{quantum computing}
\end{acronym}

%\section{I. Introduction}

\textit{Introduction.---}Extending the algorithmic toolbox and potential application areas of \ac{QC} represents an ongoing challenge for the community \cite{Montanaro2016rev,algorithms_survey}. There has been significant progress in the domain of materials science  \cite{Babbush2018,Satzinger2021,Clinton2024,ALEXEEV2024}, high-energy physics \cite{DiMeglio2024,Lejarza2024}, and chemistry \cite{McArdle2020,cao2019quantum,LeeBabbush2021,robledomoreno2024chemistry}, with potential applications in the pharmaceutical industry \cite{Santagati2024,zhou2024drugdiscovery}. Financial analysis \cite{Herman2023,mironowicz2024} and optimization \cite{Abbas2024} have also attracted attention. While in each case there are hard instances to calculate, the largest workload of high-performance computing comes from solving differential equations (computational fluid dynamics and alike), reaching exascale \cite{Heldens2021,Mullowney2021,Coti2024}. This makes scientific computing an important target area for quantum algorithmic speed-up \cite{Moller2023}. 

Motivated by the proposals for quantum linear solvers and the HHL algorithm \cite{harrow2009quantum,MartinSanz2023}, various \ac{QC}-based solvers for \acp{PDE} have been developed \cite{Berry_2014,Montanaro2016,Costa2019,Scherer2017,Krovi2023improvedquantum,Daribayev2023}, and expanded to linear combination of unitaries (LCU)-based solvers \cite{childs2017quantum,hamiltonian_simulation_lcu}, quantum signal processing (QSP)-based solvers \cite{Lin2020optimalpolynomial,Martyn2021}, adiabatic approaches \cite{subacsi2019quantum,costa2023improving,jennings2023efficient}, quantum lattice Boltzmann methods \cite{Succi2015,Steijl2023,schalkers2022efficient,schalkers2023importance,Succi_2023,Succi2024}, and quantum iterative solvers \cite{quantum_kaczmarz_method,quantum_iterative_solver,quantum_iterative_solver_2,time_marching_multigrid,continuous_quantum_jacobi}. One important feature of all these methods is that they solve a \emph{quantum} version of the linear equations problem (and corresponding \acp{PDE}), where the resulting solution is a quantum state of a $n$-qubit register. 
Reading out the full classical solution from these quantum states with tomography techniques in general scales exponentially with system size, making it prohibitive \cite{nielsen2010quantum}. Shadow tomography is an appropriate technique for some cases but for generic solutions to \acp{PDE}, the quantum states can be entangled and complicated to process with a large shadow norm \cite{Huang2020,Jerbi2024,shadow_tomography2}. This represents the readout problem for quantum linear algebra algorithms \cite{Biamonte2017}. While noted early on and discussed in the original HHL paper \cite{harrow2009quantum}, the proposed mitigation to the readout problem is in measuring \emph{some} $k$-local observable. This implies an understanding of what shall be measured. In fact, measuring certain observables can remove the exponential quantum advantage gained from inverting the \ac{PDE} \cite{Linden2022quantumvsclassical}. Extracting quantities such as principal components or high-order statistics can be done efficiently \cite{Lloyd2014} but do not always correspond to the required information. Therefore, the readout problem represents an open challenge for useful scientific computing on quantum devices. %For generic problems, shadow tomography cannot be used efficiently and so we look to build a general purpose tool for extracting information from quantum data that can be applied more widely.
%%%
\begin{figure*}[t]
    \centering
    \includegraphics[width=\linewidth]{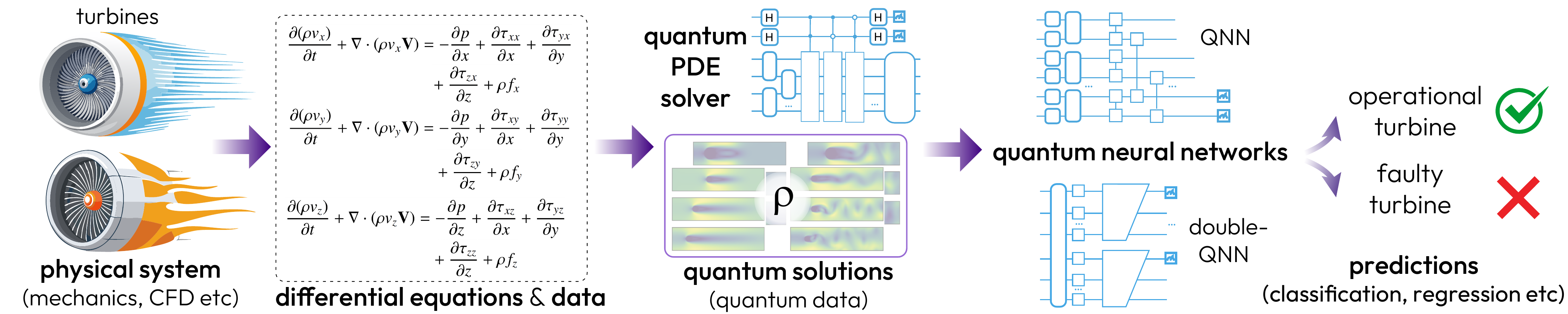}
    \caption{A quantum \ac{CFD} pipeline that uses a combination of physics-based modeling (stress, pressure, etc.) and ML to predict the behavior of an underlying system. The physics of the system is modeled by a set of \acp{PDE} that are solved with a quantum \ac{PDE} solver. The quantum solutions $\vb*{\rho}$ from the solver are then read out with a \ac{QuaSciML} model that uses a trainable observable parametrized by a \ac{QNN} to make predictions.}
    \label{fig:workflow}
\end{figure*}
%%%

\Acf{QML} aims to build tailored quantum circuits for analyzing data on quantum computers \cite{Biamonte2017,Benedetti2019rev}. Over the years the target of \ac{QML} evolved from reaching a computational advantage in analyzing Big Data \cite{Wiebe2012,Rebentrost2014PRL,Schuld2016PRA}, being a distant goal due to the classical data loading bottleneck \cite{Schuld2022PRXQ,qml2}, to the optimization-based search of optimal measurement strategies in feature spaces \cite{Schuld2019feature,Goto2021PRL,schuld2021supervised}. Based on adaptive circuit compilation, potential benefits of \ac{QML} model building include largely increased model expressivity \cite{Goto2021PRL,SchuldSweke2021PRA,Abbas2021} and quantum-native sampling that favors generative modeling \cite{Liu2018,Zoufal2019,Coyle2020,Kyriienko2024DQGM,kasture2022protocols,Rudolph2024,Wu2024}. Applied to the domain of scientific computing, quantum scientific machine learning (QuaSciML) has emerged as a subfield of \ac{QML} for solving problems based on differential equations via physics-informed modeling \cite{kyriienko2021solving,heim2021quantum,Markidis2022qpinns,Paine2021,paine2023quantum,jaderberg2023let,Jaderberg2024weather}. For models based on parametrized measurements this, by design, relaxes strict conditions for reading out the solutions and handling nonlinear terms \cite{kyriienko2021solving,Kyriienko2024DQGM}. However \ac{QML}-based modeling meets hurdles in the variational search, as trainability competes with expressivity \cite{holmes2022connecting}.  

While overcoming barren plateaus \cite{mcclean2018barren,fontana2023adjoint} remains a challenge addressed from different angles \cite{Sack2022PRXQ,pesah2021absence,larocca2023theory,puig2024warmstart,Cerezo2023CSIM}, one mode of operation has emerged as a prime beneficiary---when \ac{QML} is applied to \emph{quantum} data \cite{Huang2022,Cerezo2022qmlrev}. Quantum data represents states (density operators) prepared on quantum devices and supplied for QML-based processing (e.g. classification). This naturally favors QML in cases where shadow tomography is not applicable \cite{Cerezo2023CSIM}, and offers drastically improved generalization \cite{Huang2022,Caro2022,Caro2023,qcnn_exeter,gilfuster2023understanding,umeano2024geometric,umeano2024forrelation}. However, how to get and distribute quantum data sets from physical experiments remains a puzzle.

In this work, we propose to address the readout problem of quantum differential equation solvers by using them as a source of quantum data. We demonstrate how a \ac{QuaSciML} workflow can be adopted for classifying quantum solutions from \ac{PDE} solvers, and learning relevant features (unlike plotting full solutions). %This framework highlights how to overcome the readout problem by acquiring knowledge about the underlying physics of the modeled system without foregoing any quantum advantage achieved when using a quantum solver. 
Specifically, we show results from two \ac{CFD} problems: the classification of wave solutions obtained from the Burgers equation, and flow regimes from the Navier-Stokes equations. These results show how \ac{QuaSciML} can extract relevant information, build a problem-specific measurement operator, and highlight the influence of the chosen basis when analyzing solutions.

We note that within the field of \ac{CFD}, \ac{ML} for classification is successfully adopted, motivating the exploration of \ac{QuaSciML} capabilities. For example, \acp{CNN} were applied to enhance turbulence modeling by distinguishing between laminar and turbulent flow regimes \cite{cnn_for_turbulence, cnn_for_vorticity} as well as being used for identifying shock wave refractions in magnetohydrodynamics \cite{cnn_for_mhd}. This provides a way of automatically detecting points at which the solutions of the underlying \acp{PDE} become unsteady, improving the robustness of \ac{CFD} pipelines. Despite these benefits, \ac{ML} is limited to working with classical data (PDE solutions), and QuaSciML opens a route for analyzing large-scale \ac{CFD} problems on ever-increasing grids.

%\section{II. Quantum Scientific Machine Learning}

\textit{Quantum Scientific Machine Learning.---}\Ac{SciML} integrates data-driven models with physics-informed numerical methods to efficiently solve complex equations \cite{SciML,rackauckas2019diffeqflux,rackauckas2020universal}. Solving \acp{PDE} with \ac{SciML} requires learning a model $F:f(\vb*{x})\longmapsto\vb*{x}$ to extract a solution $f(\vb*{x})$ given data $\vb*{x}$ \cite{kyriienko2021solving,lubasch2020variational,paine2023quantum,paine2023physicsinformed}. This process can get a speed-up from quantum \ac{PDE} solvers \cite{quantum_pde_solver_theory,Lin2020optimalpolynomial,paine2023physicsinformed,quantum_iterative_solver}, where the solutions are stored as quantum states, $f(\vb*{x}_i)\rightarrow\ket{f_i}$. Thus, we arrive at the bottleneck where reading out full-grid solutions is difficult. 

Addressing the readout problem requires building an efficient model to learn from quantum data, and identifying a few relevant features from complicated states. Let us thus describe the \ac{QuaSciML} workflow for extracting information. %which seeks to combine quantum computing with \ac{SciML} techniques to extract meaningful information from quantum solutions of \acp{PDE}. \Ac{QuaSciML} leverages quantum \ac{ML} to process quantum states directly, enabling the efficient extraction of relevant physical features or informative behaviors from the model without needing full classical reconstruction. 
Consider an input space $\mathcal{X}$ representing the set of $D$ quantum solutions at a grid of points $\vb*{\rho}=\{\ket{f_i}\}_{i=1}^D$ obtained from a quantum \ac{PDE} solver and an output space $\mathcal{Y}$ wherein classification labels are mapped. Note that the solution to the \ac{PDE} $f$ is dependent on the boundary conditions and physical variables of the underlying system, which ultimately influences the class that $f$ is assigned to. %The purpose of \ac{QuaSciML} is to then perform quantum learning on this data.

\Ac{QuaSciML} models can be described formally following the idea of probabilistic concepts \cite{concept_learning}. A concept $c$ represents a function that maps input data $\vb*{x}\in\mathcal{X}$ to associated measured outcomes $y\in\mathcal{Y}$ such that $c:\mathcal{X}\longmapsto\mathcal{Y}$. Within the context of QML, a probabilistic concept represents a quantum model that accounts for the probabilistic nature of measurements. The goal is to learn from quantum data, assigning a concept to data stored in the Hilbert space such that $c_i:\ket{f_i}\longmapsto\{0,1\}\ \forall\ i \in [1,D]$. This framework is used for developing a learning concept that is based upon parameterized quantum models. The learning algorithm seeks to minimize a loss function $L_{\vb*{\theta}} = \abs{h(\vb*{\rho}, \vb*{\theta}) - c(\vb*{\rho})}^2$ to construct a hypothesis $h(\vb*{\rho}, \vb*{\theta})$, parameterized by $\vb*{\theta}$, that approximates the true concept $c(\vb*{\rho})$. The hypothesis can be developed through a variational approach or by employing a physics-informed approach that selects the optimal hypothesis from a set of measured observables representing the concept \cite{gyurik2023expsepar,umeano2024forrelation}.

The workflow of how \ac{QuaSciML} can be used within the wider context of a quantum \ac{CFD} pipeline is shown in Fig.~\ref{fig:workflow}. In terms of the quantum readout problem, the solutions (quantum states $\vb*{\rho}$) from a quantum \ac{PDE} solver can be directly passed into a \ac{QuaSciML} model for concept learning. The model then utilizes a parameterized \acf{QNN} circuit $U(\vb*{\theta})\ket{\rho}$ to extract key local and global characteristics from the quantum states. One suitable choice here is the \acf{QCNN} architecture \cite{qcnn,pesah2021absence,Herrmann2022,zapletal2023errortolerant,qfcnn,qcnn_exeter,Koki2024,sander2024qcnn}, which is trainable and enables a local readout. We also note that QCNNs can be treated in the measure-first form without variational training on QC in the adaptive way \cite{Cerezo2023CSIM,bermejo2024qcnn}, but as we work on quantum data there is still a need to identify an optimal quantum readout. Upon completing the training stage, the model learns to distinguish features such that the optimal hypothesis $h^*(\vb*{\rho}, \vb*{\theta})$ is selected to later make informative predictions with regards to the underlying system modeled by the \acp{PDE}. In terms of quantum advantage, it has been shown that \ac{QML} models can learn from exponentially fewer quantum experiments when compared with the number required by classical models \cite{qml_learning}. The goal is to establish \ac{QuaSciML} models that achieve high performance for PDE solutions at growing $n$. %These \ac{QuaSciML} models exploit the manipulation of quantum data stored in the exponentially large Hilbert space to speed-up the processing of information, enabling them to outperform their classical analogs.
%%%
\begin{figure}[t!]
    \centering
    \includegraphics[width=\linewidth]{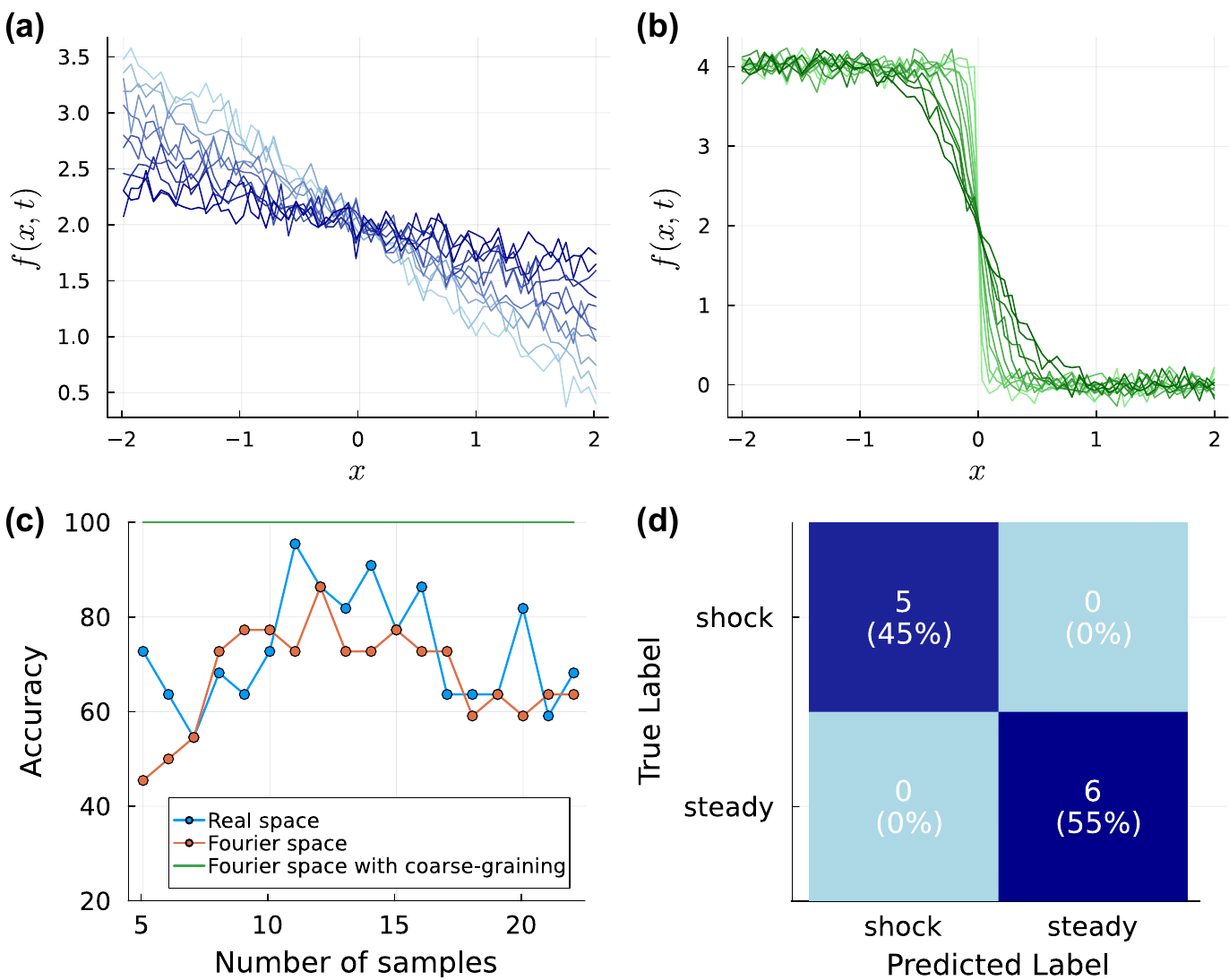}
    \caption{The classification of wave solutions with a \ac{DQNN}. \textbf{(a-b)} The one-dimensional solutions $f(x,t)$ obtained from the simulation of the Burgers equation with additional white noise. These solutions are used to classify wave behavior in terms of a steady wave (blue, left) or shock wave (green, right). Different color intensities correspond to different $t\in[0.001,590]$ snapshots. These solutions are subsequently encoded into $6$-qubit quantum states. \textbf{(c)} A comparison of the trained model accuracy as a function of the number of training samples for three \ac{DQNN} models, accounting for basis choice and coarse-graining. \textbf{(d)} The confusion matrix was evaluated on a $50\%$ test set using the optimal model that applies coarse-graining to Fourier-transformed data.}
    \label{fig:wave_solutions}
\end{figure}
%%%

The aim is to perform classification on quantum data using \acp{QNN}, optimizing a layered quantum circuit $U(\vb*{\theta})$ that informs the choice of hypothesis $h(\vb*{\rho}, \vb*{\theta})$, for constructing a sharp decision boundary. The model output is a binary label that describes \ac{CFD} simulation solutions, e.g. the type of wave detected or the category of flow behavior, without needing to directly extract information from the quantum state solutions $\ket{f}$. In particular, optimized models containing problem-specific characteristics are proposed that help to inform the decision boundary and enhance the model's accuracy through the loss minimization, $\mathrm{min}_{\vb*{\theta}}(L_{\vb*{\theta}})$.

Oftentimes, the optimal basis used for solving \acp{PDE} differs from the optimal basis used for learning from quantum data. In this sense, the quantum data $\vb*{\rho}$ obtained from a quantum \ac{PDE} solver may benefit from a basis change when uploaded to the quantum architecture used for classification. Different architectures for learning can benefit from basis changes as there can be preferred bases for performing analysis (e.g. momentum space analysis for translationally invariant solutions), influencing the ability of the model to learn the optimal concept. In composing quantum models for classification, considerations are made with regards to the basis (real space vs. Fourier space), the circuit architecture (deep-learning inspired neural networks \cite{dqnn} vs. convolutional neural networks \cite{qcnn,Caro2022,qcnn_exeter}), the introduction of a coarse-grained representation as well as common \ac{ML} characteristics like the hypothesis function and optimization strategy. Here, coarse-graining refers to concentrating on part of the solution by discarding part of a register. %taking a partial trace over the circuit. 
This reduces the complexity of states to be analyzed, helping to focus on relevant macro features (e.g. vortices or flow trends). %by tracing out certain subsystems, effectively discarding some quantum degrees of freedom and focusing on the reduced state of the remaining subsystem. 
The \ac{QFT} is used conveniently to convert the quantum data from the real space to the Fourier space \cite{Kyriienko2024DQGM,nielsen2010quantum} with an $O(n^2)$ circuit. Coarse-graining is applied afterwards to half of the register to reduce the dimensionality of the data while retaining high-frequency components of the solutions. The modified quantum data set is then fed into the model for training.

%\section{III. Results}

\textit{Results.---}This section presents the application of the \ac{QuaSciML} protocol for the purposes of binary classification. The results demonstrate the training and testing of two different quantum architectures to distinguish two fluid phenomena relevant to \ac{CFD}, namely shock wave detection and turbulence modeling. For simplicity in exhibiting how the quantum readout classification is performed, the solutions used for training and testing are obtained from classical simulations. These solutions are converted into quantum states and used as inputs to the \ac{QuaSciML} models. The results from the analysis presented here are the same as if the quantum data inputs were measured directly from a noiseless quantum \ac{PDE} solver without any classical processing. Therefore, the quantum data $\vb*{\rho}$ represent the vectors encoding the quantum state \ac{PDE} solutions in a given basis.

%\subsection{III.I. Shock wave detection}

The first problem considers \textit{shock wave detection} from solving the one-dimensional Burgers equation. This is a nonlinear \ac{PDE} that describes the propagation of waves in fluid dynamics and can be used to model various physical phenomena such as turbulence and acoustic wave propagation. The Burgers equation is given by
\begin{equation} \label{eq:burgers_equation}
    \frac{\partial f}{\partial t} + f\frac{\partial f}{\partial x} = \mu\frac{\partial^2f}{\partial x^2},
\end{equation}
where $\mu$ represents the diffusivity of the medium. For this particular problem, the goal is to classify wave solutions into either shock waves or steady waves. These solutions $f(x,t)$ are shown in Fig.~\ref{fig:wave_solutions}(a-b), where each solution represents a wave profile drawn at different snapshot times $t$. The state space is composed of $22$ solutions; $11$ steady wave solutions and $11$ shock wave solutions. These solutions are obtained from the analytical solution to the Burgers equation using similarity transformations. Additional white noise has been added to these solutions to increase the intricacy of the profiles and subsequently enhance the complexity of the learning task. 

To perform the wave behavior classification, the labeled solution set is randomly split into a training and testing set. The solutions are uploaded into quantum states using amplitude encoding, where each state is represented by $12$ qubits. The model for this problem uses a \acf{DQNN} \cite{dqnn} of depth 4 with an optimization procedure that consists of 300 gradient descent iterations at a learning rate of 0.05 and a hypothesis function given by the Pauli $Z$ measurement. The learning concept is therefore built from the hypothesis $h(\vb*{\rho}, \vb*{\theta}) = \bra{\rho}U^\dagger(\vb*{\theta})ZU(\vb*{\theta})\ket{\rho}$. The result in Fig.~\ref{fig:wave_solutions}(c) shows the accuracy of the trained model for three scenarios that consider the basis choice and the influence of coarse-graining. The number of samples refers to the number of randomly selected wave solutions drawn from the original data set for training the model, while accuracy indicates how effectively the model predicts outcomes for the entire data set.

Given the nature of this problem, it makes sense to solve the underlying differential equation in real space, as discontinuities in the solution could lead to Gibbs phenomena when using a Fourier-based approach. However, when it comes to extracting features of quantum solutions for classification, real space might not be the most suitable basis for learning. 
The optimal model uses a Fourier basis and a coarse-grained representation, which is shown to consistently achieve $100\%$ accuracy. This optimized model was then applied to the full data set when split into a $50\%$ training and $50\%$ testing set. The confusion matrix as applied to the test set is shown in Fig.~\ref{fig:wave_solutions}(d). The optimal model achieves a perfect accuracy of $100\%$, significantly outperforming both the real space model, which reaches only $27\%$, and the Fourier space model without coarse-graining, which achieves $45\%$. This highlights that bespoke trained models can enable efficient readout.

%\subsection{III.II. Turbulence modeling}

The second problem considers \textit{turbulence modeling} and the solutions of two-dimensional Navier-Stokes equations. This is a set of nonlinear \acp{PDE} that describe the motion of viscous fluid flows and can be used to model various phenomena including turbulence and aerodynamics. For this particular problem, the goal is to classify the flow around a cylinder into either the laminar or turbulent regime.  This is modeled by the Navier-Stokes equation,
\begin{equation}
    \frac{\partial\vb*{u}}{\partial t} + (\vb*{u}\cdot\vb*{\nabla})\vb*{u} = -\frac{1}{\sigma}\vb*{\nabla}p + \nu\nabla^2\vb*{u},
\end{equation}
where $\vb*{u}, \sigma, p, \nu$ respectfully represent the fluid's velocity, density, pressure, and viscosity.
%%%
\begin{figure}[t!]
    \centering
    \includegraphics[width=\linewidth]{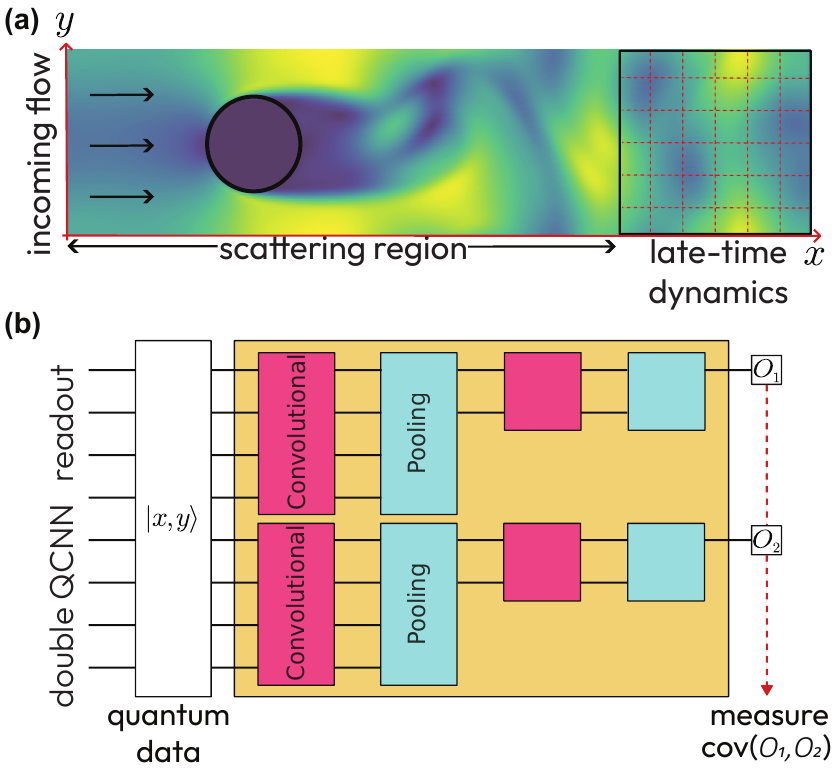}
    \caption{The encoding of quantum data into a double-\ac{QCNN} circuit for the analysis of the flow around a cylinder. \textbf{(a)} The geometrical configuration of the flow depicted in the $xy$-plane. The late-time scattering behavior of an incoming flow is subsampled from the bounded grid and encoded into the quantum state $\ket{x,y}$. \textbf{(b)} The quantum circuit showing the double-\ac{QCNN} architecture used for the classification of two-dimensional quantum data. Each \ac{QCNN} acts on a different coordinate register, parameterized by a different set of variational parameters and measured with respect to different observables $\{O_1,O_2\}$. The covariance between these observables forms the concept hypothesis and is used to learn the decision boundary.}
    \label{fig:flow_setup}
\end{figure}
%%%

The geometry of the setup is shown in Fig.~\ref{fig:flow_setup}(a), representing the flow around a cylinder. It considers the scattering of a horizontal inbound flow about a cylinder with velocity profiles modeled in the $xy$\nobreakdash-plane. The two flow classes are distinguishable by their velocity profile characteristics; for laminar flows, profiles are smooth and linear while for turbulent flows, profiles are chaotic and exhibit the von K\'{a}rm\'{a}n vortex street phenomena. These physical features are most apparent at later stages of the flow along the $x$\nobreakdash-axis, corresponding to the late-time dynamics after the flow has fully developed. To ensure that the model effectively captures these distinct dynamics, the input data is limited to the late-time behavior of the flow. This is achieved by truncating the velocity solutions to the final $25\%$ of the flow, as indicated by the bounding box in Fig.~\ref{fig:flow_setup}(a). To reduce the dimensionality of the input data and save on computational processing time, each truncated solution is further subsampled over the coarse grid. All these steps are readily implemented for quantum data by selecting relevant qubits in the quantum register and keeping track of indexing and significant bits.

The multi-dimensional nature of this problem requires encoding quantum data from grid-based inputs, necessitating that both the circuit and the model be designed to accommodate this. When uploading the solutions into quantum states with amplitude encoding, the late-time subsampled velocity arrays $\psi$ are split across two registers. The relative quantum states are structured as
\begin{equation}
    \ket{\psi} = \ket{x,y} = \sum_{i,j} \psi_{i,j} \ket{i}\otimes\ket{j},
\end{equation}
where the basis state $\ket{i}$ stores the column index corresponding to the $x$\nobreakdash-register and basis state $\ket{j}$ stores the row index corresponding to the $y$\nobreakdash-register. Then, two independent \acp{QNN} are applied; one on the top register and one on the bottom register. This circuit is named the double-\ac{QNN}, with each network having distinct trainable parameters. The quantum circuit used to implement this strategy is given in Fig.~\ref{fig:flow_setup}(b), shown specifically for QCNNs \cite{qcnn,pesah2021absence,qcnn_exeter}. 

Analyzing the correlation between the $x$\nobreakdash-register and the $y$\nobreakdash-register is an effective strategy for forming the decision boundary and devising a physics-informed hypothesis $h(\vb*{\rho}, \vb*{\theta})$. This can be measured using covariance, where laminar solutions are expected to have a low value (approaching zero), while turbulent solutions are likely to show a non-zero value. The covariance with respect to two operators $O_1$ and $O_2$ when measured across two different registers is given by 
\begin{equation}
    \text{cov}(O_1,O_2) = \tfrac{1}{2}[\langle O_1 O_2 \rangle + \langle O_2 O_1 \rangle] - \langle O_1 \rangle \langle O_2 \rangle.
\end{equation}
The operator $O_1$ is measured from the top of the first register and the operator $O_2$ is measured from the top of the second register, as shown in Fig.~\ref{fig:flow_setup}(b). A quantum model constructed with concept learning is particularly well-suited to this problem due to the architecture's inherent ability to detect entanglement, which is crucial for analyzing correlations between different registers. 
%%%
\begin{figure}[t]
    \centering
    \includegraphics[width=\linewidth]{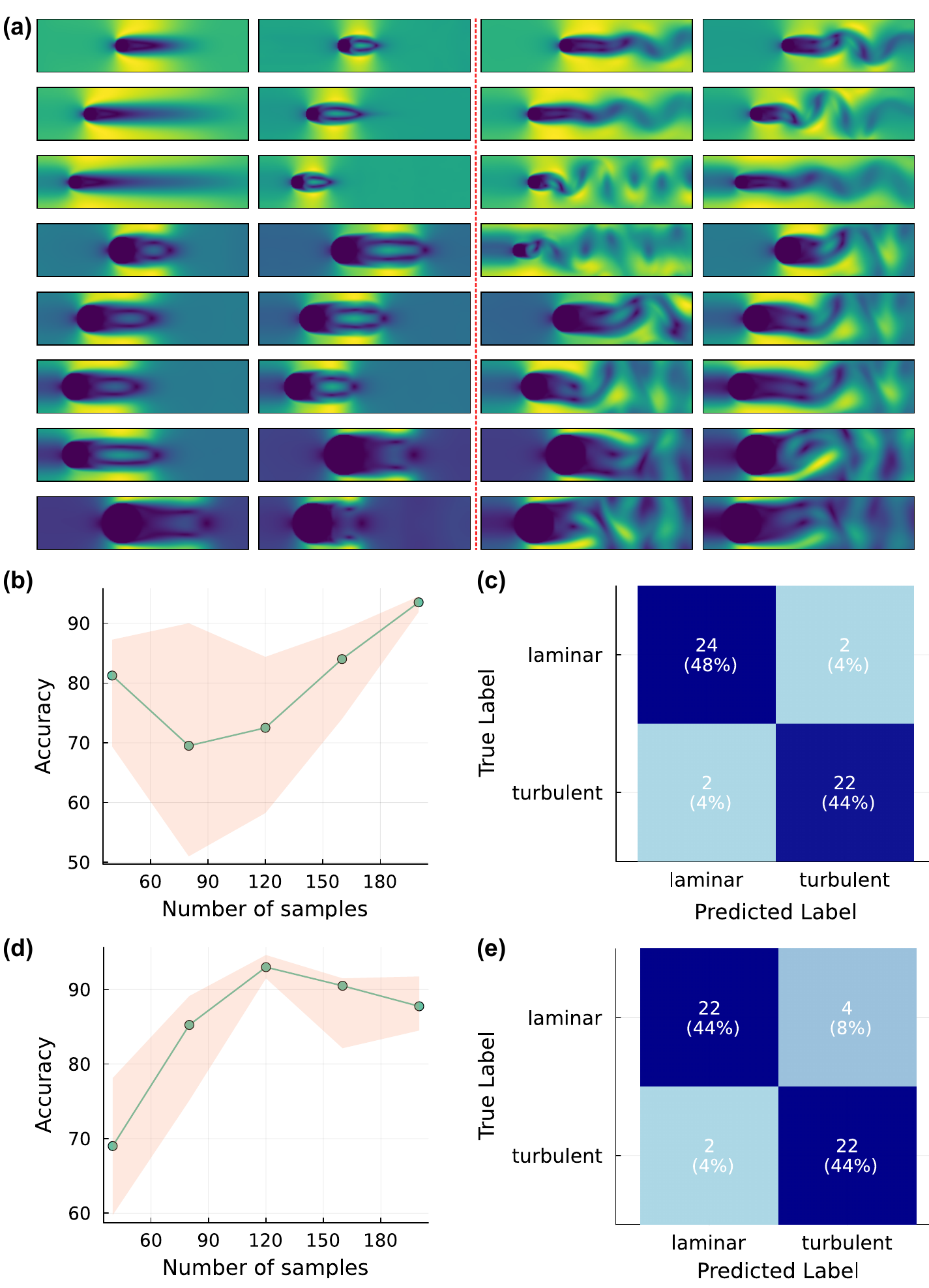}
    \caption{The classification of flow solutions with a double-\ac{QCNN}. \textbf{(a)} A subset of two-dimensional velocity solutions obtained from the simulation of the Navier-Stokes equations. These solutions are used to classify the flow in terms of a laminar regime (left-hand side) or turbulent regime (right-hand side). These solutions represent the amplitudes of corresponding quantum states encoded with $14$ qubits. \textbf{(b)} The generalization (accuracy) of the trained model as a function of the training sample size for the double-\ac{QCNN} model in the real basis. The line represents the median across four randomly chosen seeds and the shading denotes the inter-quartile range. \textbf{(c)} The confusion matrix for the double-\ac{QCNN} model in the real basis, evaluated on a $25\%$ test set as averaged across ten randomly chosen training and testing set splits. \textbf{(d)} Same as \textbf{(b)} but for the double-\ac{QCNN} model in the Fourier basis. \textbf{(e)} Same as \textbf{(c)} but for the double-\ac{QCNN} model in the Fourier basis.}
    \label{fig:flow_solutions}
\end{figure}
%%%

\par A subset of velocity solutions for different geometries and flows is shown in Fig.~\ref{fig:flow_solutions}(a). Each solution represents a snapshot of the flow profile drawn at different times. The state space is composed of $200$ solutions; $100$ laminar flow and $100$ turbulent flow solutions. These solutions are obtained with a lattice Boltzmann simulation that considers a combination of different cylinder diameters, cylinder locations along the $x$\nobreakdash-axis, and variations in the collision timescale. We note that many other approaches can be adopted and here we specifically concentrate on the analysis part rather than solving. Each complete solution presented corresponds to a quantum state using $14$ qubits. However, after truncating the late-time dynamics of the solution, the required number of qubits is reduced to $12$. Upon further subsampling over the coarse grid, the solutions are ultimately encoded onto two registers using a total of $8$ qubits. 

To perform the flow behavior classification, the labeled solution set is randomly split into $75\%$ training and $25\%$ testing sets. The solutions correspond to quantum states with amplitude-encoded solutions that are truncated and subsampled. The model for this problem considers the \ac{QCNN} \cite{qcnn} as the underlying architecture for the double-\ac{QNN}. It incorporates a hyperbolic tangent activation function $\ket{f} \rightarrow \tanh{[a(\abs{\ket{f}}-b)]} \ \forall\ \ket{f} \in \vb*{\rho}$, where the parameters $a$ and $b$ are used to shift and scale the input data appropriately. The loss function is given by the covariance of the Pauli $X$ and Pauli $Z$ operators with an optimization procedure that consists of $100$ gradient descent iterations using the parameter-shift rule at a learning rate of $0.01$. The learning concept is therefore built from the hypothesis $h(\vb*{\rho}, \vb*{\theta}) = \bra{\rho}U^\dagger(\vb*{\theta})\text{cov}(X,Z)U(\vb*{\theta})\ket{\rho}$. The result in Fig.~\ref{fig:flow_solutions}(b) shows the accuracy of the trained model using a real basis, summarized over four randomly chosen seeds. As before, the number of samples refers to the number of randomly selected velocity solutions drawn from the original data set for training the model, while accuracy indicates how effectively the model predicts outcomes for the entire data set. The average confusion matrix as applied to the test set from ten randomly chosen seeds is shown in Fig.~\ref{fig:flow_solutions}(c). The same analysis was performed on the model when evaluated in the Fourier basis. The \ac{QFT} circuit was invoked separately on the $\ket{x}$ and $\ket{y}$ registers before applying the double-\ac{QCNN} architecture. The accuracy of the trained model and average confusion matrix are shown, respectively, in Fig.~\ref{fig:flow_solutions}(d) and Fig.~\ref{fig:flow_solutions}(e). Although the real space model achieves a higher average accuracy of $92\%$ compared with the Fourier space model of $87\%$, the Fourier space model demonstrates greater stability with minimal variance in the results. The higher accuracy for the real space model makes it suitable for applications needing more precise predictions, while the Fourier space model's stability renders it more reliable in situations demanding consistent performance. In addition, the Fourier model performs better with a smaller number of samples, suggesting that the generalization and the onset of overfitting behavior differ between the two models. Overall, the optimal choice of basis for training the model will be influenced by the trade-off between accuracy, precision, and sample size.

%\section{IV. Conclusions}

\textit{Conclusions.---} We have shown that quantum scientific machine learning can be used as a tool for addressing the readout problem from quantum PDE solvers. By learning hypotheses as measurement operators from quantum data, one can bypass the need for computationally intensive post-processing of classical solutions extracted (e.g. tomographically) from quantum solvers. 
Low-dimensional output can be obtained from quantum data using models based on \acp{QNN} in the form of classification labels, representing relevant solution features, with tailored circuit architectures doing this efficiently. 

The QuaSciML tools were applied to \ac{CFD} problems of shock wave detection and turbulence modeling in scenarios where data samples come directly from quantum \ac{PDE} solvers. We find that the proposed approach can achieve perfect scores for shockwave detection and high accuracy ($>90\%$) for turbulent flow detection at high test/train split ratios. We note that the success of learning depends on problem-specific characteristics and the chosen basis. Well-crafted choices in the problem representation can also considerably reduce the required resources.

Our work opens up the area of research where \ac{QuaSciML} is inherently required as a quantum-enhanced post-processing step. 
The solutions obtained from quantum differential equation solvers will provide an abundance of quantum data (particularly, when fault-tolerant QC is available) that can be meaningfully classified. 
%A similar issue...
Extracting insights from these quantum solutions is a fundamental research question that we begin addressing here, with a long-term pursuit of efficient learning protocols. 
%This approach demonstrates the use of \ac{QuaSciML} in analyzing quantum data to learn properties of quantum differential equation solutions.

%We note that the tools and techniques used here for classification can be extended to other \ac{ML} tasks, such as clustering or regression. For example, quantum recurrent neural networks could be applied to quantum data to perform regression meanwhile, quantum Siamese neural networks could be applied to quantum data to perform detection. The latter example will require the use of a model similar to that of the double-\ac{QNN} but with the added constraint that the training parameters of both \acp{QNN} be identical. 
%Our work highlights the possibility to develop techniques for the effective readout of quantum solutions when building a functional quantum \ac{CFD} pipeline. 

\textit{Acknowledgment.---}This work was supported by Siemens Industry Software NV.

%\bibliography{bibliography}

%apsrev4-2.bst 2019-01-14 (MD) hand-edited version of apsrev4-1.bst
%Control: key (0)
%Control: author (72) initials jnrlst
%Control: editor formatted (1) identically to author
%Control: production of article title (-1) disabled
%Control: page (0) single
%Control: year (1) truncated
%Control: production of eprint (0) enabled
%

\end{document}